\documentclass[%
 reprint,
superscriptaddress,
amsmath,amssymb,
prb,
]{revtex4-2}

\usepackage{graphicx}
\usepackage{dcolumn}
\usepackage{bm}
\usepackage[colorlinks=true,linkcolor= black, citecolor = black, urlcolor = blue, plainpages=false]{hyperref}

\begin{document}

\preprint{}

\title[Ultrafast control of coherent acoustic lattice dynamics in the transition metal dichalcogenide alloy WSSe]{Ultrafast control of coherent acoustic lattice dynamics in the transition metal dichalcogenide alloy WSSe}

\author{Sergio~I.~Rey}
\email[Email: ]{s.reypuentes@amolf.nl}
\altaffiliation[current affiliation: ]{%
AMOLF, Physics of Functional Complex Matter, Science Park 102
1098 XG Amsterdam, The Netherlands}%
\affiliation{%
 DTU Electro, Technical University of Denmark, Ørsted Plads 343, 2800 Kgs. Lyngby, Denmark}%

\author{Martin~J.~Cross}%
\affiliation{DTU Electro, Technical University of Denmark, Ørsted Plads 343, 2800 Kgs. Lyngby, Denmark}%

\author{Malte~L.~Welsch}
\affiliation{%
 DTU Electro, Technical University of Denmark, Ørsted Plads 343, 2800 Kgs. Lyngby, Denmark}%

\author{Frederik~Schröder}
\affiliation{%
 DTU Electro, Technical University of Denmark, Ørsted Plads 343, 2800 Kgs. Lyngby, Denmark}%
 \affiliation{%
 NanoPhoton--Center for Nanophotonics, Technical University of Denmark, Ørsted Plads 345A, 2800 Kgs. Lyngby, Denmark}%
 
\author{Binbin~Zhou}
\affiliation{%
 DTU Electro, Technical University of Denmark, Ørsted Plads 343, 2800 Kgs. Lyngby, Denmark}%

\author{Nicolas~Stenger}
\affiliation{%
 DTU Electro, Technical University of Denmark, Ørsted Plads 343, 2800 Kgs. Lyngby, Denmark}%
 \affiliation{%
 NanoPhoton--Center for Nanophotonics, Technical University of Denmark, Ørsted Plads 345A, 2800 Kgs. Lyngby, Denmark}%
\author{Peter~U.~Jepsen}
\affiliation{%
 DTU Electro, Technical University of Denmark, Ørsted Plads 343, 2800 Kgs. Lyngby, Denmark}
\author{Edmund~J.~R.~Kelleher}
\email[Email: ]{edkel@dtu.dk}
\affiliation{%
 DTU Electro, Technical University of Denmark, Ørsted Plads 343, 2800 Kgs. Lyngby, Denmark}%
 
\date{\today}

\begin{abstract}
\noindent Coherent acoustic phonons (CAPs)—--propagating strain waves that can dynamically modify the structure and symmetry of a crystal—--offer unique opportunities for controlling material properties. 
We investigate CAP generation in the Janus-like layered alloy tungsten sulfide selenide (WS$_x$Se$_{1-x}$, hereafter WSSe). 
Employing high-fluence photoexcitation at 400~nm combined with ultrafast transient reflection spectroscopy, we capture the carrier-lattice dynamics governed by a cascade of processes including rapid exciton formation, phonon recycling, and thermoelastic deformation. 
These phenomena precede the emergence of a robust CAP mode at 27~GHz. 
Notably, the CAP amplitude in WSSe substantially exceeds that observed in the symmetric parent crystals WS$_2$ and WSe$_2$, which we attribute to an enhanced coupling mediated by a built-in out-of-plane electric field arising from the inversion asymmetry of the WSSe alloy. 
Furthermore, the implementation of a tailored two-pulse excitation sequence enables optical control of the CAP, underscoring the potential of WSSe and related Janus-like layered alloys as versatile building blocks in optomechanical and nanoacoustic device applications.
\end{abstract}

\maketitle

\section{\label{sec:intro}Introduction}
\noindent Van der Waals materials offer a promising platform for the development of future photonic, optoelectronic and hybrid devices owing to their diverse optical, electrical and thermal properties~\cite{leeMeasurementElasticProperties2008, balandinSuperiorThermalConductivity2008, bolotinUltrahighElectronMobility2008, songLargeScaleGrowth2010, soluyanovTypeIIWeylSemimetals2015, zhangTopologicalInsulatorsBi2Se32009, el-banaSuperconductivityTwodimensionalNbSe22013}. 
In particular, transition metal dichalcogenides (TMDCs)---which exhibit a thickness-dependent band gap~\cite{makAtomicallyThinMathrm2010}, high photoluminescence yield~\cite{splendianiEmergingPhotoluminescenceMonolayer2010}, and optically addressable valley selectivity~\cite{xiaoCoupledSpinValley2012, xuSpinPseudospinsLayered2014}---have attracted significant attention. 
The properties of TMDCs can be further tuned by substituting one chalcogen atom with another, producing so-called Janus monolayers.
These engineered materials not only retain the beneficial characteristics of the parent crystals but also exhibit novel attributes, including intrinsic piezoelectricity~\cite{dongLargeInPlaneVertical2017} and extended carrier lifetimes~\cite{zhengExcitonicDynamicsJanus2021}.
However, synthesizing Janus monolayers remains a significant challenge~\cite{jangGrowthTwodimensionalJanus2022, luJanusMonolayersTransition2017}. 
An alternative approach under active investigation involves the use of alloys with the general chemical formula MXY (M = Mo, W, and X, Y = S, Se, and Te). 
These alloys offer a tunable band gap and can host exotic optical states, including dark excitons and trions~\cite{puckoExcitonsTrionsWSSe2022}; however, further research is needed to determine whether they can replicate the unique properties of pristine Janus monolayers. 

Central to fully harnessing the potential of TMDCs---including their emerging Janus variants---is the ability to control their structural degrees of freedom. 
Activating coherent phonons as collective excitations can trigger phase transitions that dramatically alter materials properties~\cite{wallUltrafastChangesLattice2012}. 
In particular, using light waves to control such transitions offers promising avenues for future logic device architectures~\cite{borschLightwaveElectronicsCondensed2023}. 
Coherent acoustic phonons (CAPs), which propagate as strain waves, are useful for probing phonon-electron, phonon-plasmon, and phonon-magnon interactions~\cite{mossUltrafastStrainInducedCurrent2011, obrienUltrafastAcoustoplasmonicControl2014, kimUltrafastMagnetoacousticsNickel2012}.
With the advent of reliable ultrashort laser sources, the generation and control of GHz-THz phonon vibrations have been extensively explored~\cite{thomsenCoherentPhononGeneration1984, thomsenSurfaceGenerationDetection1986, akhmanovLaserExcitationUltrashort1992, merlinGeneratingCoherentTHz1997}. 
In TMDCs, coherent phonons have been observed in several crystal compounds, including MoS$_2$~\cite{geCoherentLongitudinalAcoustic2014}, PtSe$_2$~\cite{chenDirectObservationInterlayer2019}, PdSe$_2$~\cite{huoThicknessdependentUltrafastChargecarrier2021}, and in InSe/hBN heterostructures~\cite{greenerCoherentAcousticPhonons2018}. 
Furthermore, strain induced in layered topological insulators like Bi$_2$Se$_3$ can access novel topological phases~\cite{parkCoherentControlInterlayer2021, parkUltrafastSwitchingTopological2023a}.  
In this work, we investigate the ultrafast dynamics of Janus-like TMDC alloys---specifically WSSe---using single- and double-pump-probe spectroscopy in reflection geometry. 
Our observations are interpreted through a multi-step mechanism comprising rapid exciton formation, phonon recycling, and subsequent laser heating, which collectively generate a propagating strain wave via the thermoelastic effect.
Notably, the amplitude of the coherent oscillation in WSSe is significantly larger than that in the symmetric parent crystals WS$_2$ and WSe$_2$. 
We attribute this enhancement to the breaking of inversion symmetry by a foreign atomic species, which induces an intrinsic out-of-plane piezoelectric field that more strongly modulates the refractive index. 
This mechanism may be universal among Janus-like TMDC alloys, paving the way for novel device functionalities.

\section{\label{sec:experimental}Experimental methods}

\begin{figure*}[ht!]
    \centering
    \includegraphics[]{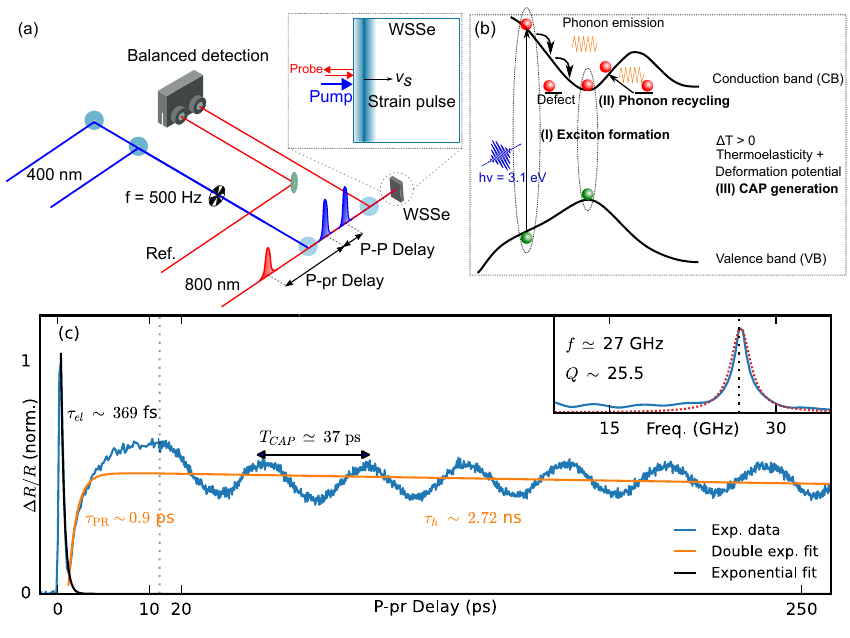}
    \caption{\textbf{Ultrafast spectroscopy of WSSe.} (a) Experimental setup. A Ti:Sapphire amplifier at 800~nm generates 100~fs pulses and is split into three arms. A weak probe and two pump pulses at 400~nm, generated by SHG in a BBO crystal, which are modulated using an optical chopper at 500 Hz. For double-pump excitation, the pump-pump (P-P) delay is kept constant. Inset: The propagating photoinduced strain pulse reflects the probe light. (b) Schematic of the photophysical processes in WSSe upon above-gap photoexcitation. (I) Thermalization and exciton formation. (II) Phonon recycling and (III) CAP generation via the deformation potential and thermoelastic mechanisms. (c) Transient reflection of WSSe. The time domain dynamics exhibit the multiple processes after photoexcitation: namely, exciton formation and thermalization, followed by a recovery of the signal caused by the phonon recycling effect, preceding the emergence of a periodic modulation due to the generation of a CAP. (d) The Fourier spectrum (blue) and its Lorentzian fit (red dots). The spectrum peaks at $\simeq$ 27~GHz and has a quality factor of $Q~\sim~25.5$. }
    \label{fig:fig1}
\end{figure*}

\subsection{Sample preparation}
High-quality TMDC crystals---WS$_2$ and WSSe (from HQ Graphene) and WSe$_2$ (from 2D semiconductors)---were studied.
Energy-dispersive X-ray (EDX) analysis of the WSSe sample revealed an atomic composition of approximately 29.8\% S and 37.0\% Se. 
The crystals were mechanically exfoliated using the scotch-tape method with 3M High-Performance Double Coated Tape 9087. 
The tape's backside was affixed to a microscope slide to support the sample, ensuring that the freshly cleaved surface (oriented parallel to the (001) plane) was accessible for reflection-mode measurements conducted under vacuum. 

\subsection{Single- and double-pump-probe spectroscopy}

Our experimental setup employs a Ti:Sapphire regenerative amplifier (Solstice Ace\textsuperscript{TM}, Spectra Physics) that generates $\sim$100~fs pulses centered at 800~nm with a 1~kHz repetition rate. 
The laser output is split into three segments: two equally intense beams serve as pump pulses, while a significantly attenuated beam is used as the probe.
To optimize the signal-to-noise ratio, the probe path remains fixed, whereas the relative timing of the pump pulses is adjusted using mechanical delay stages.
Both pump beams are frequency-doubled in $\beta$-barium borate (BBO) crystals, converting them to 400~nm pulses, while the probe remains at the original 800~nm wavelength. 
Figure~\ref{fig:fig1}(a) schematically depicts the simplified experimental layout and detection configuration. 
Both pump pulses and the probe pulse impinge on the crystal at normal incidence. 
At time $t = 0$, as shown in the inset, the pump beam generates a strain pulse (i.e. a coherent acoustic phonon, or CAP) on the surface of the WSSe sample. 
The probe is subsequently reflected by the strain pulse---that propagates at the longitudinal speed of sound, $v_s$---and by the crystal surface.    

A homodyne detection scheme is applied using a reference frequency of 500~Hz supplied by an optical chopper. 
The reflected probe signal is measured with a balanced photodiode and sequentially processed through a boxcar integrator and lock-in amplifier synchronized to the chopper frequency.
For double-pulse experiments, the chopper is strategically placed in the shared optical path of both pump beams, 
ensuring that the lock-in amplifier simultaneously captures signals originating from the influence of both pump pulses.

\section{\label{sec:cap_activation} Activation of a coherent acoustic oscillation in $\mathbf{WSSe}$}

Figure~\ref{fig:fig1}(b) illustrates the sequence of photophysical processes initiated by high-fluence excitation with 400~nm pump photons, ultimately leading to the generation of a coherent acoustic phonon. 
The transient reflection data in Fig.~\ref{fig:fig1}(c), recorded at an excitation fluence of $F\sim5$~mJ/cm$^2$, reveals these dynamics in detail. 
Initially, a single pump pulse excites carriers to high-energy states in the conduction band, driving the system out of equilibrium and producing a corresponding increase in the transient reflectivity. 
These photoexcited carriers rapidly thermalize to the bottom of the conduction band via electron-electron and electron-phonon scattering, exhibiting a single-exponential decay (black curve) with a time constant of $\tau = 369$~fs, as shown in Fig.~\ref{fig:fig1}(c). 
In addition to carrier thermalization, other competing processes occur on similar timescales.
For example, Valencia-Acuña \textit{et al}. \cite{valencia-acunaTransientAbsorptionTransition2020} demonstrated that exciton formation predominantly contributes to the fast decay in TMDC monolayers, using a photodope-pump-probe scheme with similar wavelengths. 
This interpretation is further supported by the pump photon energy of 3.1~eV, which can quasi-resonantly excite the D-exciton resonance, known to be present in both WS$_2$ \cite{cunninghamPhotoinducedBandgapRenormalization2017} and Janus monolayers of WSSe \cite{zhengExcitonicDynamicsJanus2021}. 
Similarly, other semiconductors such as GaAs \cite{vashisthaComprehensiveStudyUltrafast2021} and the ternary TMDC alloy, MoWSe, \cite{wangAnomalousDynamicsDefectAssisted2022} exhibit nonradiative trapping within the first picosecond following photoexcitation. 
At this high fluence, carrier thermalization, exciton formation, and nonradiative trapping likely compete, as not all free carriers can bind into excitons~\cite{ceballosExcitonFormationMonolayer2016, nieExcitonphononExcitonexcitonInteractions2019, nieTransientCarrierDynamics2020, trovatelloUltrafastOnsetExciton2020}. 

\begin{figure*}[ht!]
    \centering
    \includegraphics[]{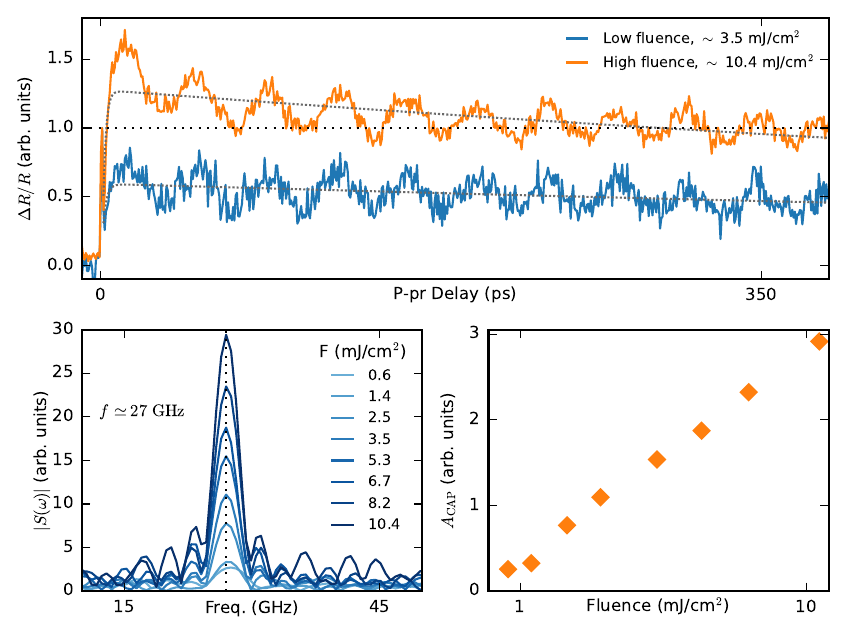}
    \caption{\textbf{Fluence dependence of CAPs.} (a) Comparison between the low- and high-fluence excitation. The curves are normalized to the exciton formation peak at short delays. For the high-fluence regime, we see that the recovered signal after the phonon recycling effect surpasses the amplitude of the peak (black dotted line). In grey, the double-exponential fits of the signal for long delays. (b) Fourier spectra as a function of fluence. The peak at 27~GHz does not exhibit a frequency shift, but the amplitude increases monotonically with fluence. (c) Amplitude of the CAP vs. fluence. The peak amplitude increases linearly with fluence. }
    \label{fig:fig2}
\end{figure*}

The transient reflection signal fully decays within 1~ps and subsequently recovers over the remainder of the experimental delay range. 
We fit the subsequent rise and decay of the transient reflection with a double-exponential model, highlighted by the orange line. 
The rising component has a time constant of $\tau_r = 950$~fs and is attributed to the phonon recycling (PR) mechanism. 
In this process, free carriers thermalize to the bottom of the conduction band while emitting phonons that result in an increased lattice temperature~\cite{wangAnomalousDynamicsDefectAssisted2022}. 
These incoherently generated phonons can transfer energy to carriers trapped in midgap states, repopulating the conduction band.
This effect has been observed across various systems---including symmetric alloys (MoWS$_2$), asymmetric alloys (MoSSe and WSSe), and binary TMDCs (MoS$_2$, WS$_2$)~\cite{wangAnomalousDynamicsDefectAssisted2022}---and is analogous to the mechanism reported in graphene-WS$_2$ heterostructures, where acoustic phonon energy is transferred from graphene to WS$_2$~\cite{weiAcousticPhononRecycling2020}. 
The slower decay component, characterized by a time constant of $\tau_h = $~2.72~ns, is ascribed to heat dissipation via thermal diffusion. 
Similar nanosecond-scale decays due to thermal diffusion have been observed in WTe$_2$~\cite{soranzioUltrafastBroadbandOptical2019} and other multilayered semiconductors~\cite{xiaoNanosecondMidinfraredPulse2019}, under comparable pump-probe conditions. 
While various long-lived processes, such as phonon bottlenecks~\cite{chiObservationHotphononEffect2020}, intervalley exciton scattering~\cite{wagnerTrapInducedLong2021}, defect-bound excitons~\cite{liuVisualizingHotCarrierExpansion2022}, and Auger recombination~\cite{adhikariGeneratingCapturingSecondary2022} have been reported in TMDC monolayers, none extend into the nanosecond regime.

\begin{figure}[ht!]
    \centering
    \includegraphics[width=1\linewidth]{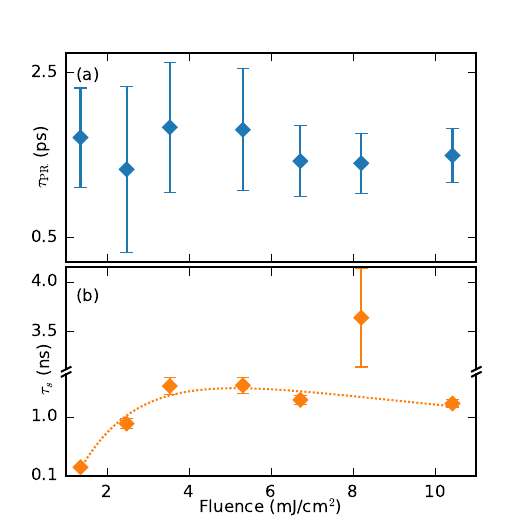}
    \caption{\textbf{Characteristic time constants in WSSe.} Constant $\tau_\mathrm{PR}$ as a function of fluence (a) and double-exponential behavior of $\tau_\mathrm{h}$, fitted with orange dotted line (b).}
    \label{fig:fig2_1}
\end{figure}

The CAP manifests as a periodic modulation superimposed on the slower exponential decay of the transient reflection signal. 
After subtracting the double-exponential background and applying a fast Fourier transform (FFT), we observe a prominent oscillation at approximately 27~GHz (see the power spectrum in the inset of Fig.~\ref{fig:fig1}(c)), corresponding to a period of $T_{CAP} = 37~\mathrm{ps}$. 
This modulation arises from the constructive and destructive interference between the probe light reflected from the crystal surface and that reflected from the propagating CAP---a process discussed further in Section~\ref{sec:comparison}. 
We propose that the observed longitudinal CAP is primarily generated via the deformation potential (DP) mechanism. 
In this scenario, photoexcitation alters the carrier density, which in turn modifies the overlap of atomic orbitals, thereby generating phonons and inducing a differential volumetric change~\cite{ruelloPhysicalMechanismsCoherent2015}. 
Since this process occurs during the laser pulse---a timescale much shorter than the phonon lifetime---the resulting phonons are generated simultaneously and propagate as coherent wavepackets with a uniform phase. 
Furthermore, we rule out the possibility of transverse CAP generation, as such modes cannot be detected when both the pump and probe beams impinge on the sample at normal incidence~\cite{matsudaCoherentShearPhonon2004}.   

The DP mechanism is widely regarded as the primary source of CAP generation in semiconductor materials. 
Although thermoelastic deformations become increasingly dominant at high fluences~\cite{ruelloPhysicalMechanismsCoherent2015, youngPicosecondStrainPulses2012}, we propose that, under our experimental conditions---characterized by high fluence and high-energy photon excitation---the DP mechanism remains dominant.
This hypothesis is supported by the observation that the carrier lifetime (on the nanosecond scale) is much longer than the CAP oscillation period of $\sim37$ ps~\cite{maUltrafastGenerationDetection2023}. 
Moreover, the increase in lattice temperature due to carrier thermalization primarily serves to sustain carriers in the conduction band via the PR mechanism, rather than directly deforming the crystal.
In this context, re-excitation of trapped carriers leads to phonon annihilation, ultimately depleting the phonon population~\cite{wangAnomalousDynamicsDefectAssisted2022}. 

The quality factor $Q$ serves as a key figure of merit for comparing oscillatory systems~\cite{priyaPerspectivesHighfrequencyNanomechanics2023, greenerCoherentAcousticPhonons2018}. 
It is defined as the ratio of the oscillation frequency relative to the spectral width of the oscillation (related to the oscillation lifetime in the time-domain): $Q = f/\Delta f$. 
Due to limitations in our experimental delay range, we approximate the lifetime by fitting the oscillatory component with a decaying cosine function of the form $ A \exp{(-t/\tau_{osc})} \cos{(2\phi f t + \phi)}$,
where the decay constant, $\tau_{osc} = 355$ ps.
We then extend the time axis based on this decay constant and calculate the FFT, as shown in the inset of Fig.~\ref{fig:fig1}(c). 
By fitting a Lorentzian profile to the Fourier peak, we extract the spectral width $\Delta f$ and thus estimate $Q$. 
Our analysis yields $Q~\sim~25.5$. 
While this value is lower than the highest Q reported in optomechanical and nanophononic platforms~\cite{priyaPerspectivesHighfrequencyNanomechanics2023}, it is notably higher than the quality factors observed in hBN~\cite{greenerCoherentAcousticPhonons2018} and acoustoplasmonic nanocrosses~\cite{obrienUltrafastAcoustoplasmonicControl2014}.    

We can vary the thermoelastic contribution to the coherent oscillation by increasing the pump fluence. 
Figure~\ref{fig:fig2}(a) compares transient reflection signals at two different fluences, normalized to the sharp peak at zero delay (indicated by the horizontal dotted black line). 
At a fluence of $F~\sim~10.4$~mJ/cm$^2$, the recovered signal exceeds the initial peak. 
Such behavior cannot be explained solely by the PR mechanism---even at 100\% efficiency, re-exciting all trapped carriers would produce a recovered signal no greater than the original peak, given that its magnitude is proportional to the number of photoexcited carriers. 
Therefore, we infer that the amplitude of the coherent oscillation increases with fluence because thermoelastic deformation becomes more dominant. 
Figure~\ref{fig:fig2}(b) shows the power spectra of the oscillatory component after subtracting the non-oscillatory background. 
Although no frequency shift is observed with increasing fluence, the amplitude of the CAP increases linearly, as shown in Fig.~\ref{fig:fig2}(c) with orange diamonds, with no detectable sign of saturation. 
This linear dependence supports the conclusion that thermoelasticity increasingly governs the CAP generation at high fluences~\cite{geCoherentLongitudinalAcoustic2014, youngPicosecondStrainPulses2012}.

We show the evolution of the rising and decaying exponential components as a function of fluence in Figs.~\ref{fig:fig2_1}(a) and (b), respectively. 
The fast rise time, which we attribute to the PR mechanism, remains nearly constant with fluence, indicating that it is dominated by the intrinsic phonon dynamics. 
This observation is consistent with previous studies~\cite{wangAnomalousDynamicsDefectAssisted2022}. 
In contrast, the slow component---fitted with a double-exponential curve---initially increases with fluence, suggesting that heat diffusion becomes less efficient as the phonon population grows. 
For fluences above 5~mJ/cm$^2$, however, the time constant decreases, as enhanced phonon-phonon scattering becomes more probable~\cite{wangAnomalousDynamicsDefectAssisted2022, prasadNonequilibriumPhononDynamics2023, chiUltrafastCarrierPhonon2019}. 
Notably, the data point at a fluence of $\sim8~\mathrm{mJ/cm^2}$ deviates from the general trend; this discrepancy could be resolved by extending the scanning range. 
Furthermore, we observe no change in the fast decay constant at short delays, which reinforces the conclusion that exciton formation is the dominant process at early times~\cite{valencia-acunaTransientAbsorptionTransition2020}.     

\begin{figure*}[ht!]
    \centering
    \includegraphics[]{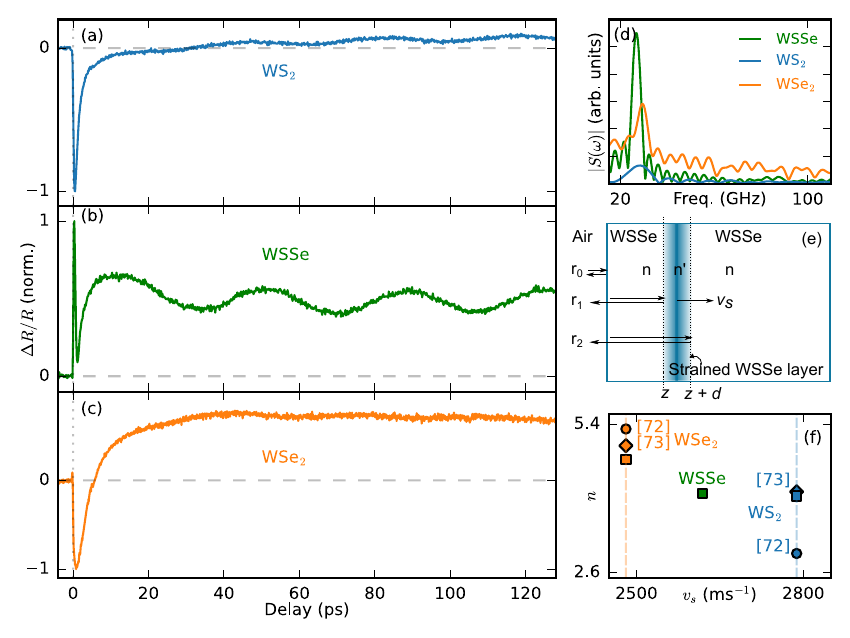}
    \caption{\textbf{Coherent acoustic phonons in the symmetric parent TMDCs.} (a)-(c) Transient reflection of tungsten-based TMDCs at a fluence of~$\sim$~5~mJ/cm$^2$. While both WS$_2$ (blue) and WSe$_2$ (orange) exhibit a similar periodic modulation, their amplitudes are drastically reduced compared to WSSe (green). (d) The Fourier spectrum for each sample peaks at a slightly different frequency. (e) Schematic view of a propagating strained WSSe layer. The pump pulse generates a longitudinal coherent acoustic phonon that travels at the speed of sound $v_s$. The relevant reflection processes in the Fabry-Perot cavity, represented by the arrows, are used to calculate the reflection from the sample. The index of refraction in the strained layer, $n'$, is perturbed with respect to the rest of the sample due to the strong strain-induced piezoelectric field. This model can also be applied to WS$_2$ and WSe$_2$. (f) Comparison of refractive indices and speed of sound in WS$_2$ (blue), WSSe (green), and WSe$_2$ (orange). Different markers denote literature values together with our own, as described in the main text. }
    \label{fig:fig3}
\end{figure*}

\section{\label{sec:comparison} Comparison with $\mathbf{WS_2}$ and $\mathbf{WSe_2}$}
To elucidate the influence of crystal asymmetry on the transient reflection dynamics, we compared the responses of symmetric TMDCs with that of the Janus-like alloy WSSe. 
Figures~\ref{fig:fig3}(a), (b) and (c) display the normalized transient reflection signals for WS$_2$, WSSe, and WSe$_2$, respectively, under similar experimental conditions. 
The transient reflection dynamics of both parent TMDCs show a negative behavior shortly after photoexcitation. 
In semiconducting materials, this is typically attributed to photoinduced absorption, as additional states created by the pump pulse become available in the conduction band. 
In WS$_2$, the transient reflection characteristics are well described by a double exponential function, where the fast component reflects rapid thermalization (and potentially exciton formation)~\cite{valencia-acunaTransientAbsorptionTransition2020, ceballosExcitonFormationMonolayer2016}, while the slower component corresponds to carrier recombination processes~\cite{vashisthaComprehensiveStudyUltrafast2021}. 
Notably, the sign of the transient reflection signal is inverted after a characteristic delay---approximately 35~ps for WS$_2$ and 10~ps for WSe$_2$. 
This inversion is likely a result of renormalization of the bandgap, where a high density of electrons induces a change in the gap energy~\cite{vashisthaComprehensiveStudyUltrafast2021, cunninghamPhotoinducedBandgapRenormalization2017, fukudaCoherentOpticalResponse2024, chernikovPopulationInversionGiant2015}. 
The differences in inversion times may be linked to their distinct bandgaps-- about 1.2~eV for bulk WS$_2$ and 1.35~eV for WSe$_2$~\cite{kamDetailedPhotocurrentSpectroscopy1982, gusakovaElectronicPropertiesBulk2017}---as well as other factors that require more detailed microscopic modeling. 
Additionally, the WSe$_2$ signal exhibits an extra rapid increase in $\Delta R / R$ immediately after the arrival of the pump pulse, commonly attributed to a hole-filling effect observed in some semiconductors~\cite{maUltrafastGenerationDetection2023}. 

The transient reflection signals of all three materials display periodic modulation superimposed onto the overall response; notably, however, the amplitude of the oscillatory component in WSSe is significantly larger than in the symmetric TMDCs. 
Figure~\ref{fig:fig3}(d) shows the power spectra of the normalized signals after subtraction of the non-oscillatory background. 
Background removal can introduce artifacts in the low frequency region ($\leq 10$~GHz)~\cite{greenerCoherentAcousticPhonons2018}, and thus we have omitted this portion of the spectrum for clarity. 
Table~\ref{tab:freq_Q} summarizes the oscillation frequencies and the quality factors, $Q$, for the three TMDCs, determined using the method described in Section~\ref{sec:cap_activation}. 

\begin{table}[ht!]
    \centering
    \begin{tabular}{|c|c|c|} \hline
        TMDC &  $f$ (GHz) & $Q$  \\  \hline \hline 
        WS$_2$  & 28.2 & 4.9 \\ \hline
        WSe$_2$ & 29.4  & 1.4 \\ \hline
        WSSe &  26.9 & 25.5 \\ \hline
    \end{tabular}
    \caption{Frequency and quality factor of the CAP in the three TMDC crystals studied.}
    \label{tab:freq_Q}
\end{table}
Analysis of the Fourier spectra and the quality factors clearly indicates that the CAP activated in the WSSe alloy exhibits both a higher amplitude and a longer lifetime than those observed in the symmetric parent TMDCs. 
This distinct behavior of the acoustic phonon dynamics is intrinsic to the alloy rather than representing an intermediate state between its parent materials. 
To explain the enhanced CAP, we propose that piezoelectricity plays a critical role. 
The substitution of sulfur with selenium disrupts the inversion symmetry of the crystal, thereby enabling the development of a piezoelectric field under mechanical strain~\cite{gautschiPiezoelectricSensorics2002}. 
Thermodynamically, the induced lattice distortion---caused by the replacement atom---increases the configurational entropy of the material~\cite{yehNanostructuredHighEntropyAlloys2004}, which has been suggested as a strategy to enhance piezoelectric properties~\cite{chenSolutionbasedFabricationHighentropy2021}. 
This mechanism has been demonstrated in alloyed nitrides~\cite{tasnadiOriginAnomalousPiezoelectric2010, talleyImplicationsHeterostructuralAlloying2018}, in ternary TMDC alloys~\cite{chen2DTransitionMetal2022}, and has been predicted for MXY alloys~\cite{yuMechanicalElasticityPiezoelectricity2019} as well as in perfectly asymmetric Janus monolayers~\cite{dongLargeInPlaneVertical2017}. 
Additionally, strong electron-phonon interactions in piezoelectric resonators can markedly reduce energy loss mechanisms~\cite{gokhalePhononElectronInteractionsPiezoelectric2014}, which may account for the superior figure of merit of the WSSe alloy observed in our experiments.     

\begin{figure*}[ht!]
    \centering
    \includegraphics[]{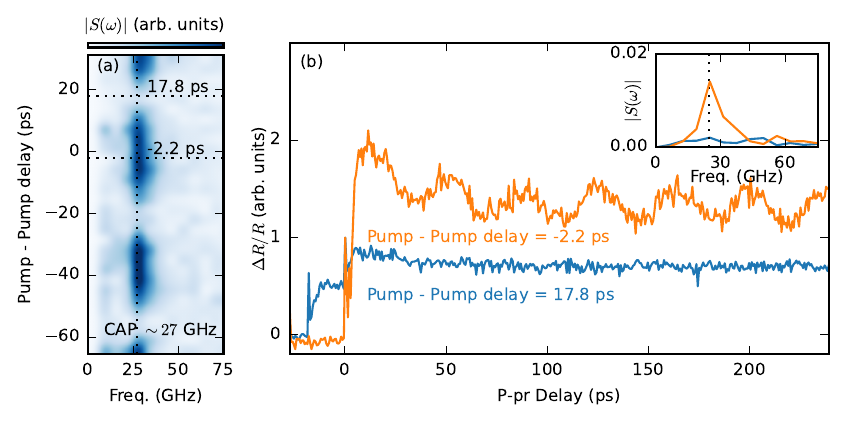}
    \caption{\textbf{All-optical control of CAPs in WSSe.} (a) 2D map of frequency vs. pump-pump delay dynamics. 
    (b) Time domain dynamics for in-phase and out-of-phase excitation with pump pulses of similar fluence. 
    The selected values for pump-pump delays are highlighted in the 2D map with black dotted lines. 
    The curves are normalized to the initial peak of the static pump. 
    For out-of-phase pump pulses (blue) the periodic modulation is completely quenched while the in-phase (orange) excitation shows an enhanced modulation. 
    Inset: the comparison between Fourier spectra in both cases shows that the peak of the out-of-phase excitation has decreased below the noise level.}
    \label{fig:fig4}
\end{figure*}

To describe the CAP, we adopt the framework proposed by Liu~\textit{et al.} in reference~\cite{liuFemtosecondPumpprobeSpectroscopy2005} for GaN/InGaN piezoelectric heterostructures. 
Given our photoexcitation and detection geometries, we assume that only longitudinal strain pulses are activated---thus, only the out-of-plane elastic constant, $C_{33}$, is required. 
A schematic of the macroscopic model is shown in Fig.~\ref{fig:fig3}(e). 
The photoinduced strain pulse is represented as a thin, uniformly strained layer in the sample, where the refractive index $n'$ deviates slightly from that of the unstrained material, $n$. This strained layer is located at a depth $z = v_s \tau$, where $\tau$ denotes the pump-probe delay, and its thickness $d$ is treated as a phenomenological parameter corresponding to the spatial extent of the CAP wavepacket. 
The speed of sound $v_s$ is given by~\cite{liuFemtosecondPumpprobeSpectroscopy2005}:
\begin{equation}
    \label{eq:vs}
    v_s = \sqrt{C_{33}/\rho}.
\end{equation}
where $\rho$ is the mass density.
We further assume that the modified refractive index remains constant within the strained region, i.e. for $z \in [z,z+d]$. 
The transient reflection signal is then calculated by summing the first three terms of the Fabry-Perot series, $r = r_0 + r_1 + r_2$, yielding~\cite{liuFemtosecondPumpprobeSpectroscopy2005}:

\begin{equation}
    \label{eq:model}
    \frac{\Delta R}{R}(\omega, t) \propto |\delta n|\sin{(kd)}\sin{\left( 2\pi \frac{t}{T} + \phi \right)}, 
\end{equation}

\noindent where $\phi$ is the phase of the CAP and $k$ the magnitude of the wavevector. 
The oscillation period, $T$, is given by the relation~\cite{thomsenSurfaceGenerationDetection1986, cardonaModulationSpectroscopySemiconductors1970}:
\begin{equation}
    \label{eq:period}
    T = 1/f = \frac{\lambda}{2 v_s n}, 
\end{equation}
\noindent with $\lambda$ denoting the probe wavelength. 
A key outcome of this model is that the amplitude of the coherent oscillation depends exclusively on the change in the refractive index and the thickness of the strained layer, while the oscillation frequency is determined by the longitudinal speed of sound and the refractive index at the probing wavelength. 
The enhanced oscillation amplitude observed in WSSe suggests that the strain-induced modulation of the refractive index is significantly stronger in this piezoelectric material, likely due to the combined effects of the induced piezoelectric field and the Franz-Keldysh effect~\cite{sunCoherentAcousticPhonon2000, liuFemtosecondPumpprobeSpectroscopy2005}.

Due to the absence of experimental measurements for the optical constants of WSSe alloys, we estimate its refractive index at 800~nm using Eq.~\eqref{eq:period}. 
First, we calculate the speed of sound from Eq.~\eqref{eq:vs}. 
To do so, we approximate the mass density by combining the known sulfur-to-selenium ratio in the alloy with the reported densities of WS$_2$~\cite{eaglesonmaryConciseEncyclopediaChemistry2011} and WSe$_2$~\cite{agarwalGrowthConditionsCrystal1979}. 
We adopt a value of $C_{33} = 5.53 \times 10^{10}$ $\mathrm{Pa}$ for WSSe multilayers from density functional theory (DFT) simulations~\cite{dongLargeInPlaneVertical2017}. 
With these parameters, the speed of sound is calculated as $v_s = \sqrt{C_{33} / \rho} = 2618.2 \text{ m/s}$, which, when applied to Eq.~\ref{eq:period} yields a refractive index of $ n = 4.09$. 
Figure~\ref{fig:fig3}(f) compares our calculated refractive index for WSSe with experimental values for the parent TMDCs. 
The x-axis shows the longitudinal speed of sound for bulk WSe$_2$ (orange) and WS$_2$ (blue), as determined by Muratore \textit{et al.}~\cite{muratoreCrossplaneThermalProperties2013}. 
The y-axis displays refractive indices measured for three-layer samples by Hsu \textit{et al.}~\cite{hsuThicknessDependentRefractiveIndex2019} (circles) and for bulk samples via ellipsometry by Munkhbat \textit{et al.}~\cite{munkhbatOpticalConstantsSeveral2022} (diamonds).   
The square markers represent our own calculation of the refractive indices using Eq.~\eqref{eq:period} and the corresponding coherent phonon frequencies from table \ref{tab:freq_Q}. 
Our estimated value for WSSe shows strong agreement with the bulk experimental data from Ref.~\cite{munkhbatOpticalConstantsSeveral2022}, although it is slightly lower than that of WS$_2$. This difference is consistent with Eq.~\eqref{eq:period}, given the lower oscillation frequency observed in WSSe. 
Nevertheless, the refractive index of WSSe is expected to lie within the limits of the range defined by its parent compounds.
It is reasonable that the DFT-derived Young's modulus for an ideal Janus crystal~\cite{dongLargeInPlaneVertical2017} differs from the effective modulus in our experimental WSSe alloy, which may account for the modestly lower refractive index estimated by our measurements.  

\section{All-optical control of the coherent acoustic oscillation}
We record the two-dimensional dynamics by sequentially acquiring a series of one-dimensional pump-probe traces, each incorporating an additional pump pulse precisely delayed relative to the original excitation using a mechanical delay stage. 
The second time axis corresponds to the delay introduced by the second pump pulse. 
The fluence of both pump pulses is F~$\sim$~5~mJ/cm$^2$, ensuring that they generate coherent acoustic oscillations of comparable amplitude. 
Figure~\ref{fig:fig4}(a) presents a 2D map of frequency versus pump-pump (P-P) delay. 
In our experimental convention, a negative pump-pump delay indicates that the second pump pulse trails the first, whereas a positive delay indicates that it leads the first. 
For example, a delay of -2.2~ps corresponds to a trailing second pump pulse, while a delay of 17.8~ps corresponds to a leading second pump pulse.
In the 2D map, the CAP at 27~GHz is highlighted by a vertical black dotted line, together with two horizontal line cuts (highlighted again by black dotted lines) at these representative delays. 
Figure~\ref{fig:fig4}(b) shows the corresponding time-domain transient reflection signals ($\Delta R/R$) for the corresponding pump pulse sequences, normalized to the fast exciton peak of the original pump pulse (at zero delay). 
The orange curve, recorded at a pump-pump delay of -2.2~ps, exhibits an enhanced oscillatory component, indicating that the CAP wavepackets generated by the two pump pulses constructively interfere.
In contrast, the blue curve, obtained at a pump-pump delay of 17.8~ps, shows the characteristic recovery of the transient reflectivity after photoexcitation attributed to the PR mechanism, but a complete suppression of the periodic modulation, evidencing destructive interference of coherent phonon wavepackets. 
The inset, which displays the corresponding spectral magnitude of both time-domain traces, further confirms that the CAP is quenched under out-of-phase photoexcitation. 

\begin{figure}[t!]
    \centering
    \includegraphics[width=1\linewidth]{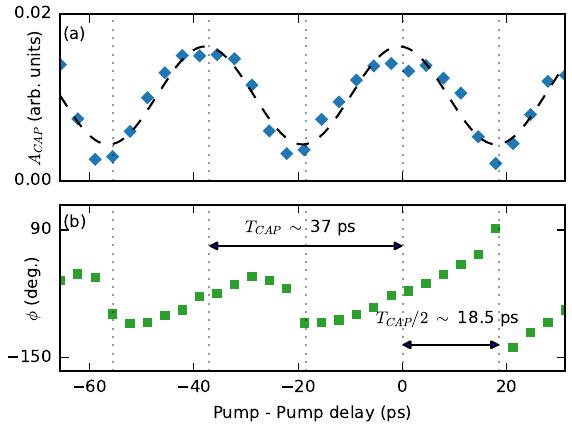}
    \caption{\textbf{All-optical control of CAPs in WSSe.} (a) Complete P-P delay dependence of the phonon peak. The amplitude varies as a cosine of the modulation frequency, which represents constructive and destructive interference between coherent acoustic phonon wavepackets. (b) P-P delay dependence of the Fourier phase. Abrupt changes might indicate a back-reflected strain pulse.}
    \label{fig:fig5}
\end{figure}
 
A general rule is that when the pump-pump delay $\simeq~(2n~+~1)T_{CAP}/2$, the CAPs interfere destructively, whereas delays near an integer multiple of $T_{CAP}~=~37$~ps lead to constructive interference. 
This behavior is illustrated in Fig.~\ref{fig:fig5}(a), where the magnitude of the CAP at 27~GHz is plotted as a function of the delay between pump pulses, following the black vertical line highlighted in Fig.~\ref{fig:fig5}(a). 
The experimental data (blue diamonds) follows a periodic function described by $\cos{(2\pi f t + \varphi)}$, as shown by the black dashed line. 
In our analysis, we set the phase constant $\varphi$ to zero, assuming that at zero pump-pump delay the pulses are perfectly in phase, yielding maximal constructive interference and the highest oscillation amplitude.  

Figure~\ref{fig:fig5}(b) depicts the spectral phase as a function of the pump-pump delay, with vertical dotted lines provided as guides. 
Starting from complete destructive interference at a pump-pump delay of -18.5~ps, the phase of the oscillating signal exhibits an approximately linear dependence on the delay. 
After one period---at a pump-pump delay of 18.5~ps---the phase exhibits an abrupt jump of $\Delta \phi = 245^{\circ}$. 
This phase jump recurs each time the amplitude reaches a minimum due to destructive interference. 
A similar phenomenon has been observed in InGaN/GaN quantum well structures~\cite{sunCoherentOpticalControl2001}, where the shift is attributed to a reversal of the propagation direction of the acoustic wave. 
Further investigation, however, is needed to fully understand this behavior in bulk WSSe, where the crystal is assumed to be effectively infinite along the propagation direction. 

In conclusion, we investigated the time- and frequency-domain dynamics of bulk WSSe under high-fluence, high-energy photoexcitation using single- and double-pump-probe spectroscopy in reflection geometry. 
Our experiments reveal a sequence of photophysical processes with distinct time constants.
At short delays, an initial fast exponential decay---attributable to rapid carrier thermalization, nonradiative trapping, and exciton formation under quasi-resonant excitation of the D-exciton---is observed.
This is followed by a recovery of the transient reflection signal with a characteristic time constant of 950~fs, which we attribute to repopulation of conduction band states via a phonon recycling mechanism.
At longer delays, a coherent acoustic phonon (CAP) emerges, generated primarily by the deformation potential mechanism and enhanced by thermoelastic effects at higher fluences. 
Notably, the amplitude and lifetime of the coherent oscillation in WSSe are significantly greater than those in WS$_2$ and WSe$_2$, suggesting that an intrinsic out-of-plane piezoelectric field in the asymmetric WSSe crystal more effectively modulates the refractive index than in the symmetric TMDCs. 
Moreover, by employing an ultrafast double-pump excitation scheme, we confirmed the coherent nature of the generated acoustic phonons through the optical control of their amplitude and phase. 
By temporally separating the pump pulses on a timescale comparable to the phonon oscillation period, we achieved complete quenching for out-of-phase excitation and enhancement for in-phase excitation.
The cosine dependence of the amplitude of the coherent oscillation as a function of pump-pump delay is a signature of interference between the coherent phonon wavepackets. 
These findings underscore the potential of asymmetric TMDCs for high-performance actuators and nanoresonators and motivate further research into synthesizing ideal Janus multilayered materials.
Additionally, our two-pulse experiments demonstrate the feasibility of an all-optical approach for ultrafast control of a material's structural degrees of freedom. 

\begin{acknowledgments}
S.I.R. and E.J.R.K. acknowledge financial support from the Independent Research Fund Denmark Sapere Aude grant (project number 9064-00072B). 
F.S. and N.S. acknowledge support from the Danish National Research Foundation through NanoPhoton--Center for Nanophotonics (grant number DNRF147). 
N.S. acknowledges funding by the Novo Nordisk Foundation NERD Programme (project QuDec NNF23OC0082957). 
\end{acknowledgments}

\section*{Author declarations}

\subsection*{Conflict of interest}

The authors declare no conflict of interest. \\

\subsection*{Author contributions}

\textbf{Sergio I. Rey:} conceptualization (lead); data curation (lead); formal analysis (lead); investigation (lead); methodology - setup design and sample preparation (lead); validation (equal); visualization (lead); writing - original draft (lead); writing - review and editing (equal). 
\textbf{Martin J. Cross:} data curation (supporting); methodology (supporting); validation (equal); writing - review and editing (supporting).
\textbf{Malte L. Welsch:} data curation (supporting); methodology (supporting); validation (equal); writing - review and editing (supporting).
\textbf{Frederik Schröder:} methodology - sample preparation (equal); validation (equal); writing - review and editing (equal).
\textbf{Binbin Zhou:} methodology (supporting); supervision (equal); validation (equal); writing - review and editing (equal).
\textbf{Wiebke Albrecht:} supervision (supporting); validation (supporting); writing - review and editing (supporting).
\textbf{Nicolas Stenger:} conceptualization (equal); methodology - sample preparation (supporting); validation (equal); writing - review and editing (equal).
\textbf{Peter U. Jepsen:} funding acquisition (equal); investigation (equal); methodology (equal); resources (equal); supervision (equal); validation (equal); writing - review and editing (equal). 
\textbf{Edmund J. R. Kelleher:} conceptualization (equal); funding acquisition (lead); investigation (equal); methodology (equal); resources (lead); supervision (lead); validation (equal); writing - review and editing (lead).  

\section*{Data Availability Statement}

The data that support the findings of this study are available from the corresponding author upon reasonable request.

\section*{References}
\bibliography{clean_bib}

\end{document}